\documentclass[twocolumn,twoside]{revtex4}
\usepackage{graphicx}
\usepackage{fancyhdr}
\usepackage{multirow}
\usepackage{color}

\begin{document}

\title{B lifetime and $B^0-\bar B^0$ mixing results from early Belle II data}

%

\author{J. Kandra, T. Bilka \\
on behalf of the Belle II Collaboration}
\affiliation{Charles University, Prague, Czech republic}

\begin{abstract}
\noindent
The Belle II experiment at the SuperKEKB energy-asymmetric $e^+ e^-$ collider is a substantial upgrade of the B factory facility at the Japanese KEK laboratory. The design luminosity of the machine is $8\times 10^{35}$ cm$^{-2}$s$^{-1}$ and the Belle II experiment aims to record 50 ab$^{-1}$ of data, a factor of 50 more than its predecessor. From February to July 2018, the machine has completed a commissioning run, achieved a peak luminosity of $5.5\times 10^{33}$ cm$^{-2}$s$^{-1}$, and Belle II has recorded a data sample of about 0.5 fb$^{-1}$. Main operation of SuperKEKB has started in March 2019. We use this dataset to characterize the performance of the detector regarding the tracking of charged particles, the reconstruction of known resonances, and the capability of identifying displaced decay vertices. To assess the B Physics capabilities of the experiment, one of the first benchmarks consists in the measurement of the lifetime of B mesons and of the $B^0-\bar B^0$ mixing frequency. We present the first results, based on samples of B mesons that decay to hadronic and semileptonic final states.
\end{abstract}

\maketitle

\thispagestyle{fancy}


\section{Introduction}
\noindent
The Belle II detector is built on asymmetric electron-positron accelerator, SuperKEKB, in Tsukuba, Japan. The acceleraor is designed to deliver more than 55 billions of $\mathrm{B\bar{B}}$ pairs in next ten years. Many of upgrades was done at the detector. The vertex detector, is located in centre of the Belle II detector, was installed to providing precise measurement of vertex position produced particles \cite{DesignReport}. It is composed of silicon semiconductor sensors organised to six layers. Two inner layers are build by DEPFET based pixel sensors and double sided strip sensors are mounted to four outer layers. After installation in December 2018 calibration and determination of alignment corrections of vertex detector was necessary to realize. One is used alignment procedure based on a minimization algorithm. The alignment corrections and occurrence of systematic misalignments should be validated. Systematic misalignments of vertex detector is introduced here. A tool, is able to clarify or verify geometry of the vertex detector, is described. There are presented Monte Carlo studies focusing to proving correct functionality of the tool and first experiences using early Belle II data. The semileptonic techniques for B lifetime and $B^0-\bar B^0$ measurement are shown. 

\begin{figure}[h]
\centering
\includegraphics[width=80mm, trim={50mm, 0mm, 50mm, 0mm}, clip]{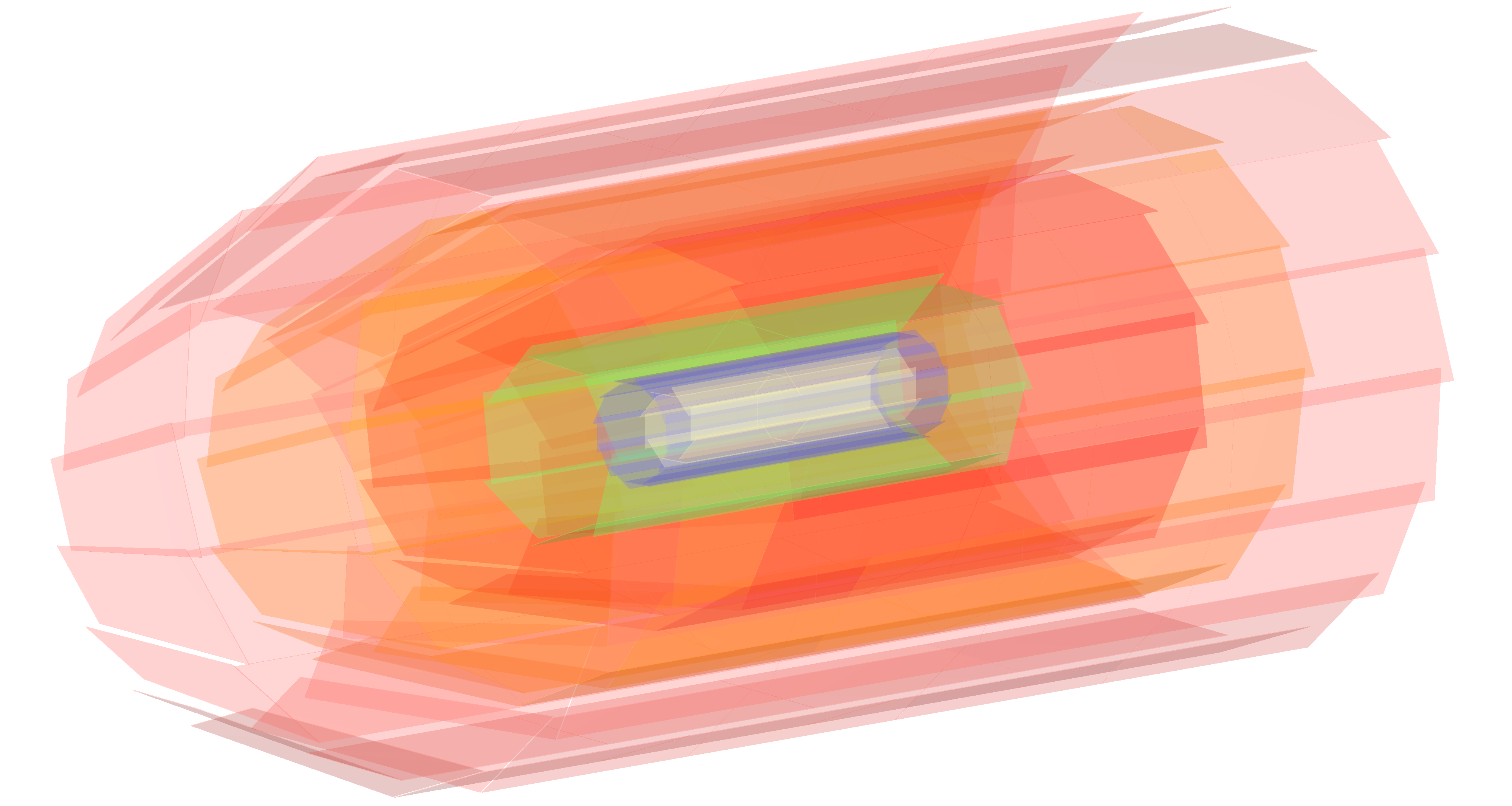}
\caption{The Belle II vertex detector} \label{fig:ImpactToPhysics}
\end{figure}

\section{Impact of $\chi^2$ invariant modes to physics}
\noindent
The track based alignment procedure of the Belle II vertex detector is based on minimization algorithm \cite{KandraThesis}. Optimally minimization algorithm will determine real alignment corrections. Otherwise algorithm can converge to local minimums are called $\chi^2$ invariant modes or weak modes. The weak modes are difficult recognized by standard monitoring and validation methods, but their effect to physics is significant. \\

\begin{figure}[h]
\centering
\begin{tabular}{c|c|c|c}
 & \emph{$\Delta r$} & \emph{$r \Delta \phi$} & \emph{$\Delta z$} \tabularnewline 
 \hline
 \rule{0pt}{2ex}    
 \multirow {5}{*}{$r$} & \textbf{Radial exp.} & \textbf{Curl} & \textbf{Telescope} \tabularnewline
 & $\Delta r = c_{s} \cdot r$ & $r \Delta \phi = c_{s} \cdot r + c_0$ & $\Delta z = c_{s} \cdot r$  \tabularnewline
 & \includegraphics[height=11mm]{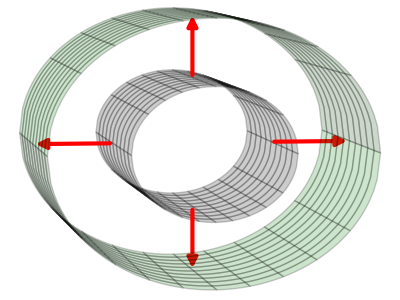}  & \includegraphics[height=11mm]{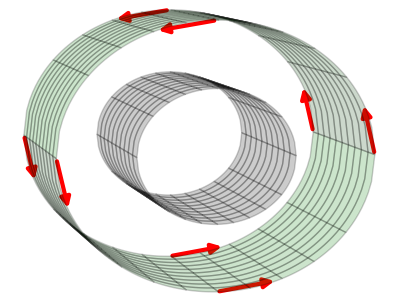}  & \includegraphics[height=11mm]{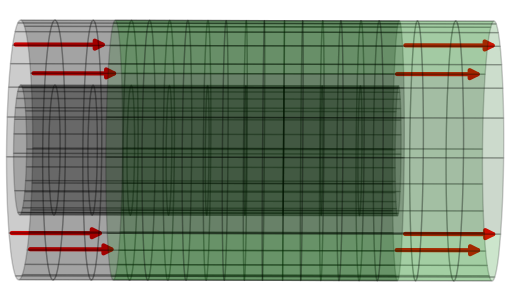} \tabularnewline
 \hline
 \rule{0pt}{2ex} 
 \multirow {5}{*}{$\phi$} & \textbf{Elliptical exp.} & \textbf{Clamshell} & \textbf{Skew} \tabularnewline
 & $\Delta r = c_{s} \cdot \cos{(2\phi)} \cdot r$ & $\Delta \phi = c_{s} \cdot \cos{(\phi)}$ & $\Delta z = c_{s} \cdot \cos{(\phi)}$  \tabularnewline
 & \includegraphics[height=11mm]{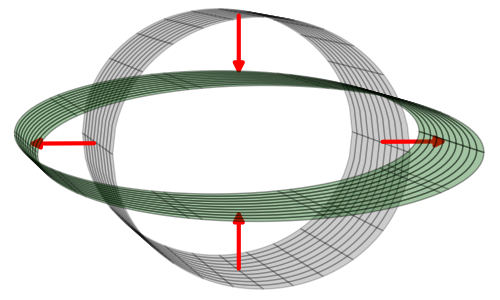}  & \includegraphics[height=11mm]{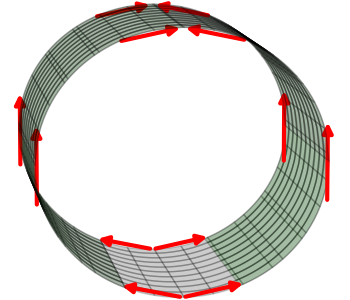}  & \includegraphics[height=11mm]{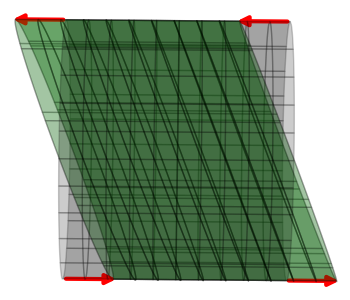} \tabularnewline
 \hline 
 \rule{0pt}{2ex} 
 \multirow {5}{*}{$z$} & \textbf{Bowing} & \textbf{Twist} & \textbf{Z exp.} \tabularnewline
 & $\Delta r = c_{s} \cdot |z|$ & $r \Delta \phi = c_{s} \cdot z$ & $\Delta z = c_{s} \cdot z$  \tabularnewline
 & \includegraphics[height=11mm]{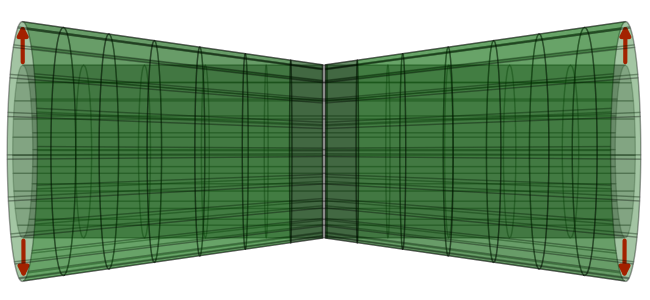}  & \includegraphics[height=11mm]{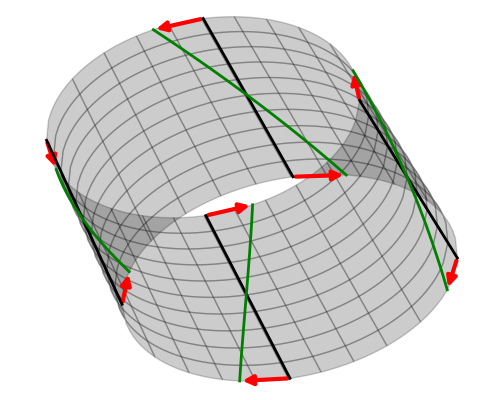}  & 
\includegraphics[height=11mm]{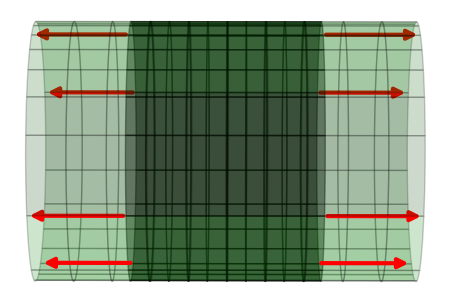} \tabularnewline
\end{tabular}
\caption{Weak modes of the vertex detector}
\label{fig:WeakModes}
\end{figure}

\noindent
The definition of $\chi^2$ invariant modes can be expressed using standard tracking parametrization in cylindrical coordinate system \cite{FPCP2017}. The classification of weak modes with parametrization can be found in figure~\ref{fig:WeakModes}.

\noindent
Time dependent CP violation measurement is used for describing impact of $\chi^2$ invariant modes to physics. Our Monte Carlo study (figure~\ref{fig:ImpactToPhysics}) is based on measuring of $A_{CP}$ and $S_{CP}$ variables as function of systematic misalignment of vertex detector. Systematic misalignments were modelled using coherent movements of sensors with maximal value about 250 $\mathrm{\mu m}$. 
\begin{figure}[h]
\centering
\includegraphics[width=80mm]{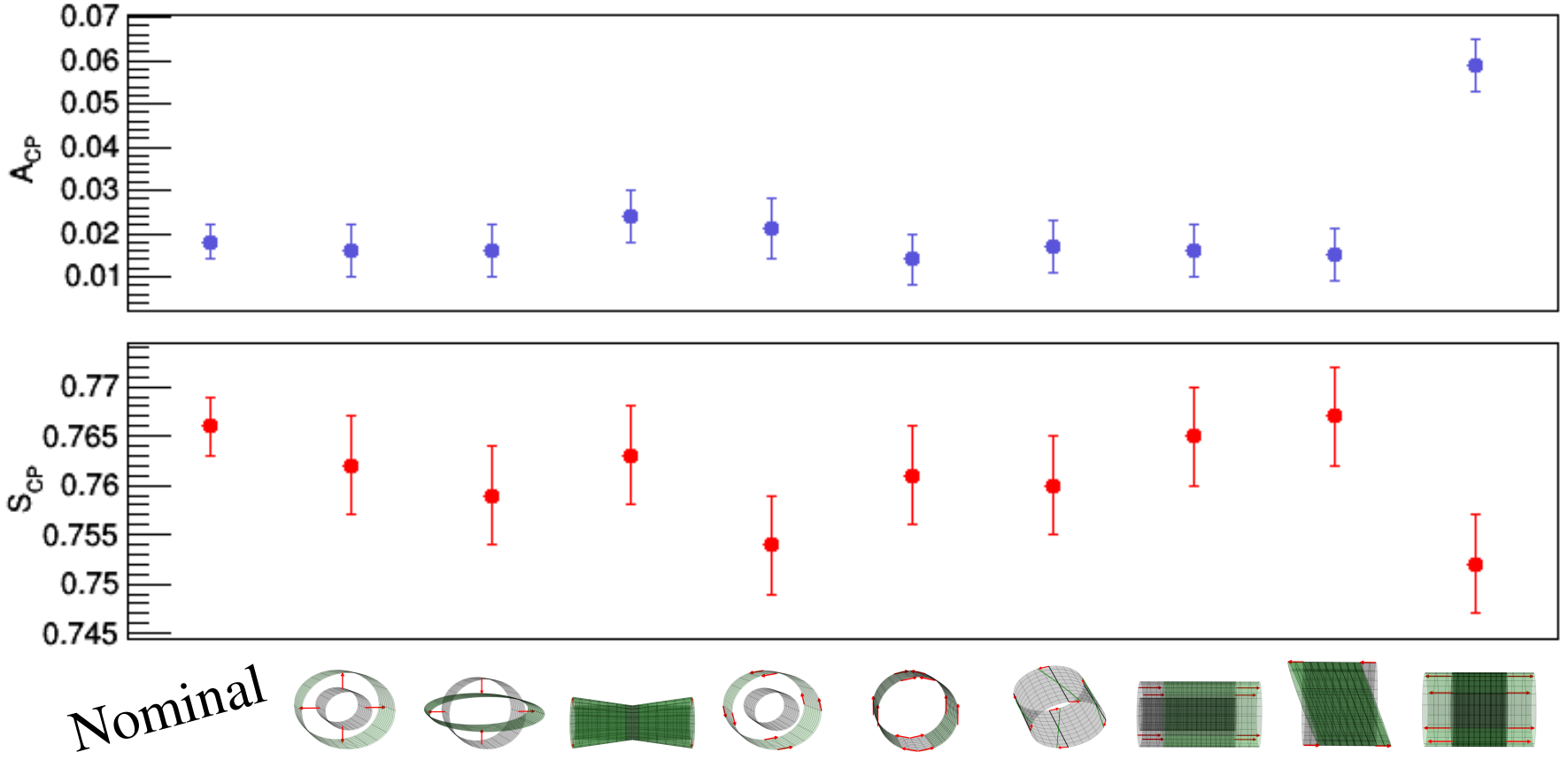}
\caption{Impact of weak modes to CP violation measurement: Significat effect to physics is seen for Z expansion of vertex detector.} \label{fig:ImpactToPhysics}
\end{figure}

\section{Monitoring $\chi^2$ invariant modes}

\noindent
The problematic of systematic misalignments of silicon detector is known very well from previous experiments (e.g. LHC experiments \cite{LHCAlignment}). The several ways of monitoring weak modes, from physical to pure tracking techniques, can be used. In our studies we were looked for a technique with universal application with possibility to determine several weak modes, optimally all weak modes. From known techniques we select pure tracking technique based on overlap residuals \cite{HeinemannThesis}.

\begin{figure}[h]
\centering
\includegraphics[width=80mm, trim={0mm, 135mm, 0mm, 75mm}, clip]{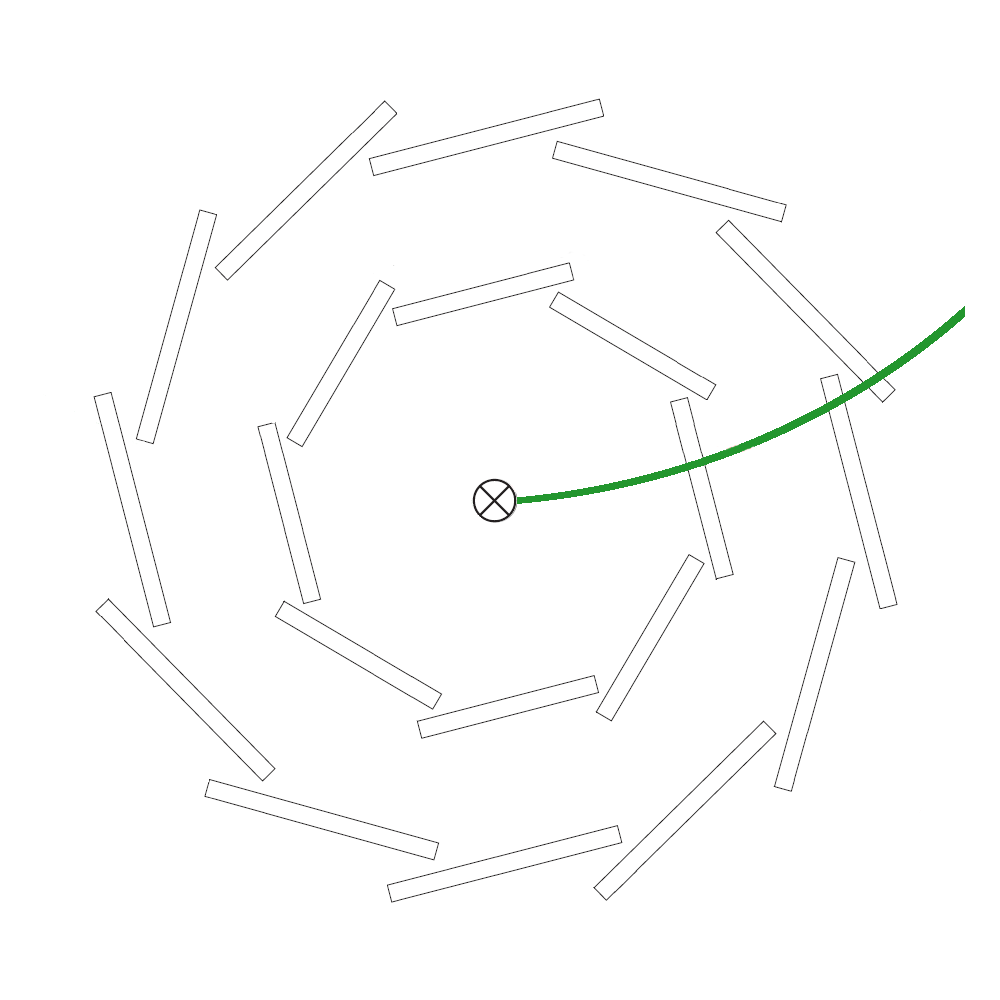}
\caption{A tracks with overlapping hits in schematic view.} 
\label{fig:SchematicView}
\end{figure}

\begin{figure}[h]
\centering
\includegraphics[width=45mm, trim={160mm, 280mm, 120mm, 0mm}, clip]{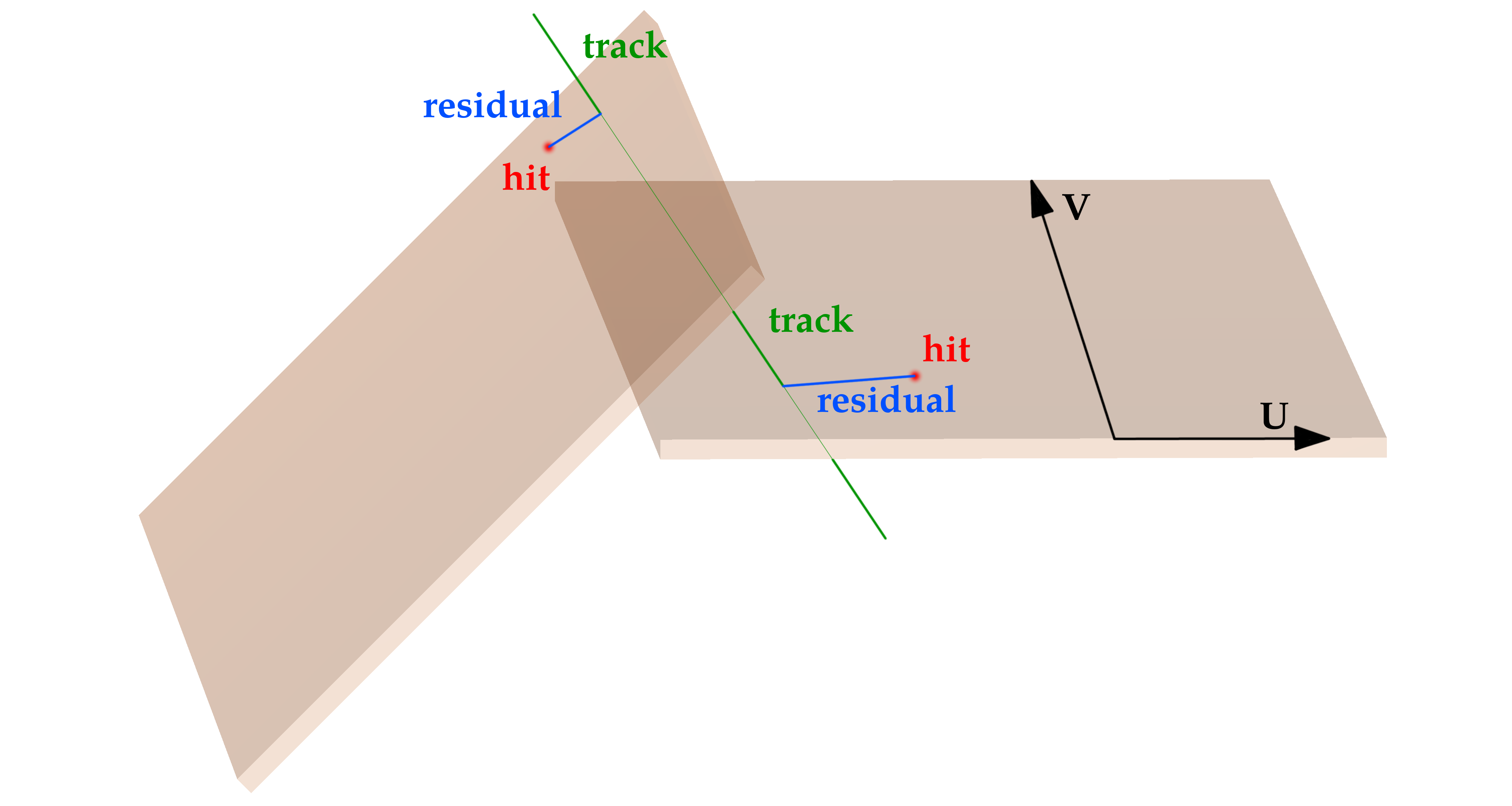}
\includegraphics[width=55mm, trim={0mm, 0mm, 0mm, 0mm}, clip]{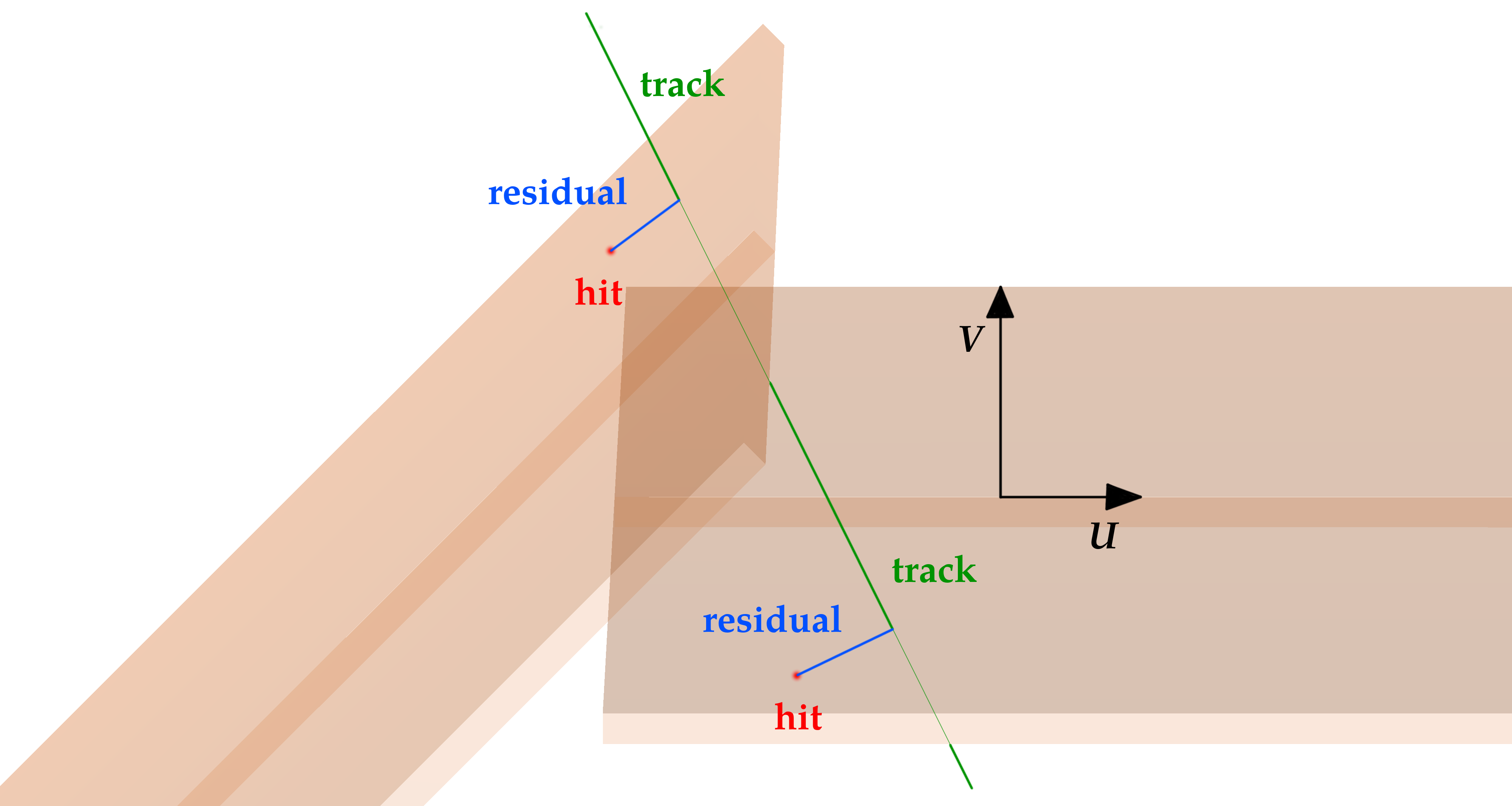}
\caption{Classification of overlapping hits: hits with same (top) and different (bottom) position of sensors in ladders} 
\label{fig:Clasification}
\end{figure}

\noindent
The technique is based on searching a track with hits in overlapping area of layers, which are built from ladders of sensors (figure~\ref{fig:SchematicView}). From selected tracks overlapping hits are separated from standard hits and both groups are analysed. From our studies we are able to identified two different types of overlapping hits: hits with same and different position of sensors in ladders (figure~\ref{fig:Clasification}). Both types have different properties and statistical occurrence. The overlapping hits with same position of sensors in ladders are sensitive on geometrical effects in $r$ and $\phi$ directions with 100-times higher occurrence than hits with different position of sensors in ladders, which are able to catch effects in $z$ direction.

\noindent
Useful variable for monitoring is residual difference of overlapping hits, where residual is defined as difference between measured and expected position of hit. The pixel and strip sensors used in the Belle II vertex detector are able provide 2D measurement on each sensor. Because of that we are able to obtain 2D residual difference for a pair of overlapping hits. Residual differences are measured for both collections of overlapping hits. Monitoring tool is collecting hits from non-overlapping area too for additional knowledge about detector geometry.   

\section{Monte Carlo studies and datasets}
\begin{figure}[h]
\centering
\begin{tabular}{c|c} 
\textbf{Nominal} & \textbf{Clamshell} \tabularnewline
 \includegraphics[width=35mm, trim={144mm, 0mm, 144mm, 8mm}, clip]{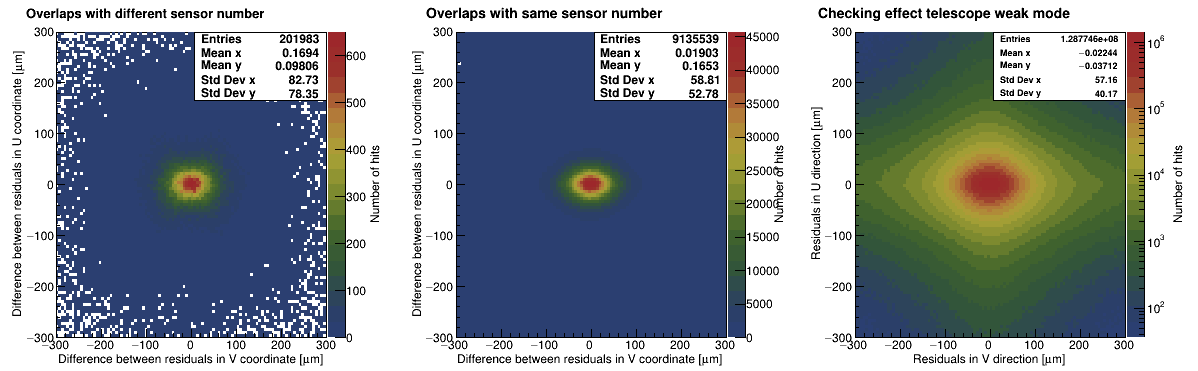}    &
 \includegraphics[width=35mm, trim={144mm, 0mm, 144mm, 8mm}, clip]{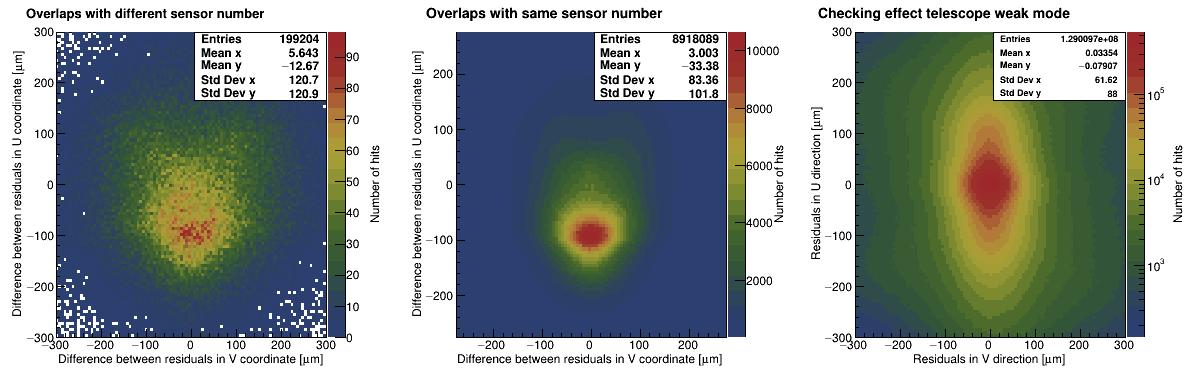} \tabularnewline \hline
\textbf{Radial expansion} & \textbf{Twist} \tabularnewline
 \includegraphics[width=35mm, trim={144mm, 0mm, 144mm, 8mm}, clip]{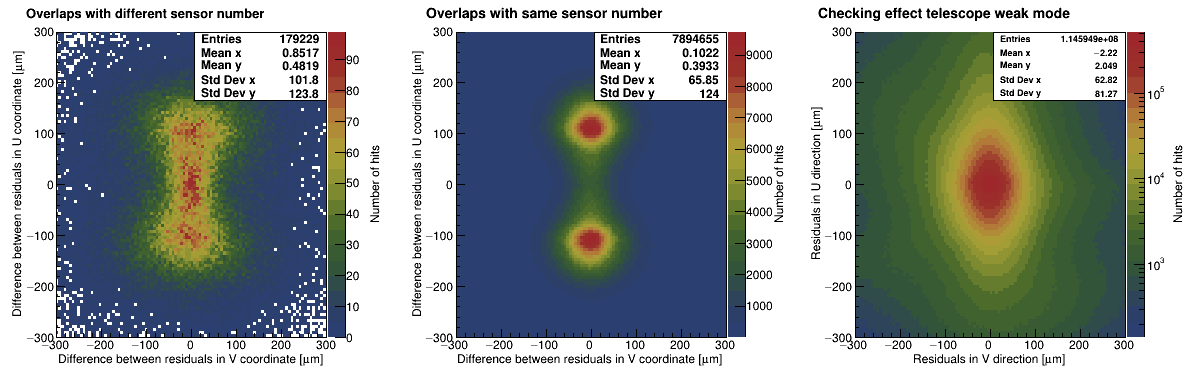}    &
 \includegraphics[width=35mm, trim={144mm, 0mm, 144mm, 8mm}, clip]{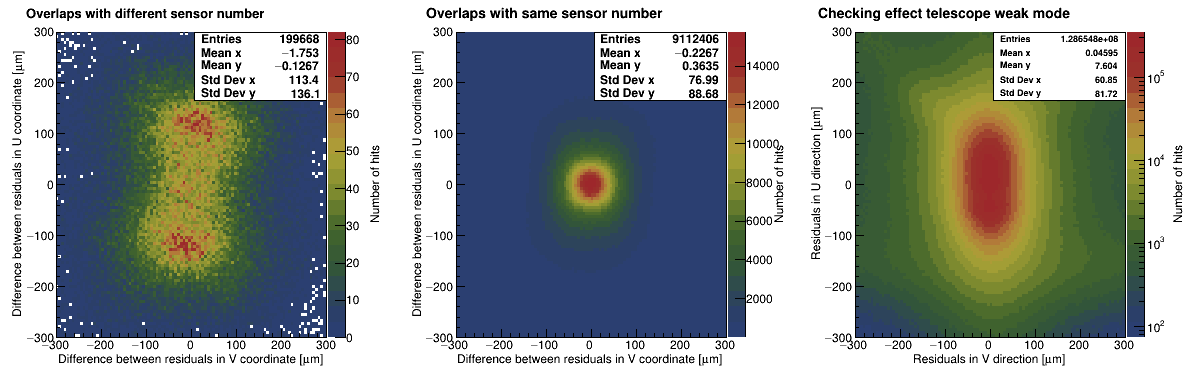} \tabularnewline \hline
\textbf{Elliptical expansion} & \textbf{Telescope} \tabularnewline
 \includegraphics[width=35mm, trim={144mm, 0mm, 144mm, 8mm}, clip]{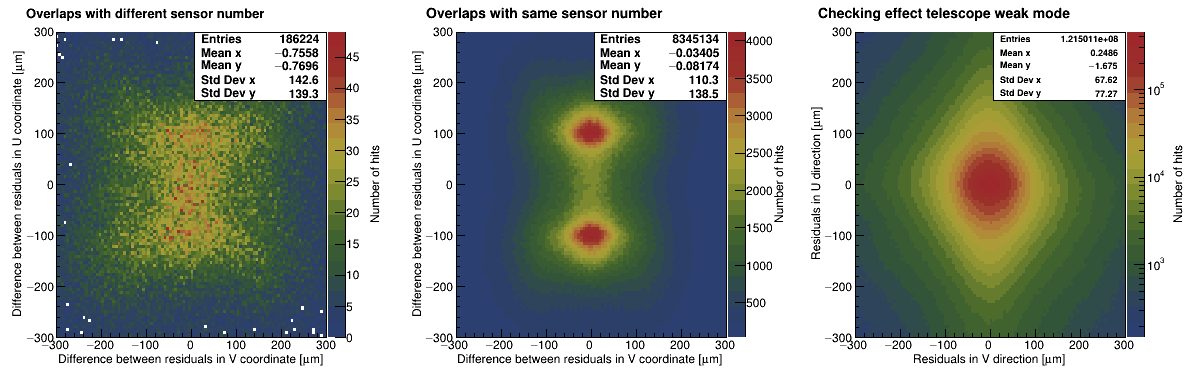}    &
 \includegraphics[width=35mm, trim={144mm, 0mm, 144mm, 8mm}, clip]{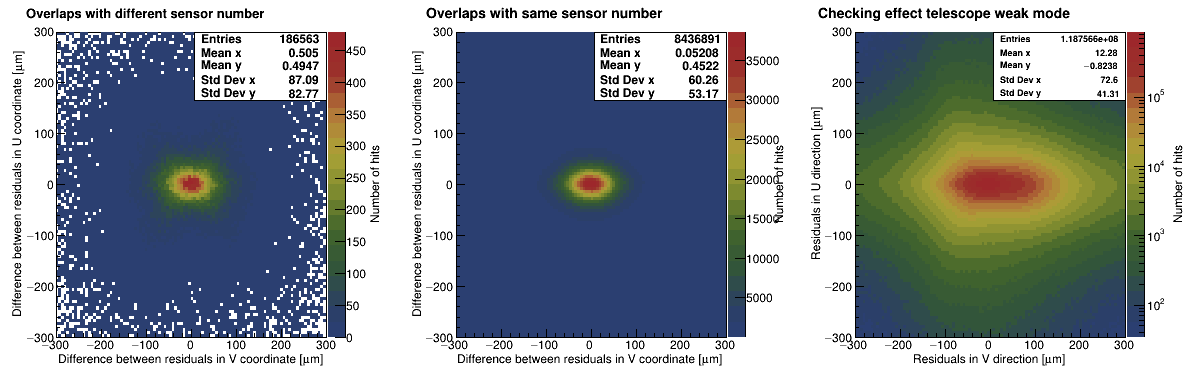} \tabularnewline \hline
\textbf{Bowing} & \textbf{Skew} \tabularnewline
 \includegraphics[width=35mm, trim={144mm, 0mm, 144mm, 8mm}, clip]{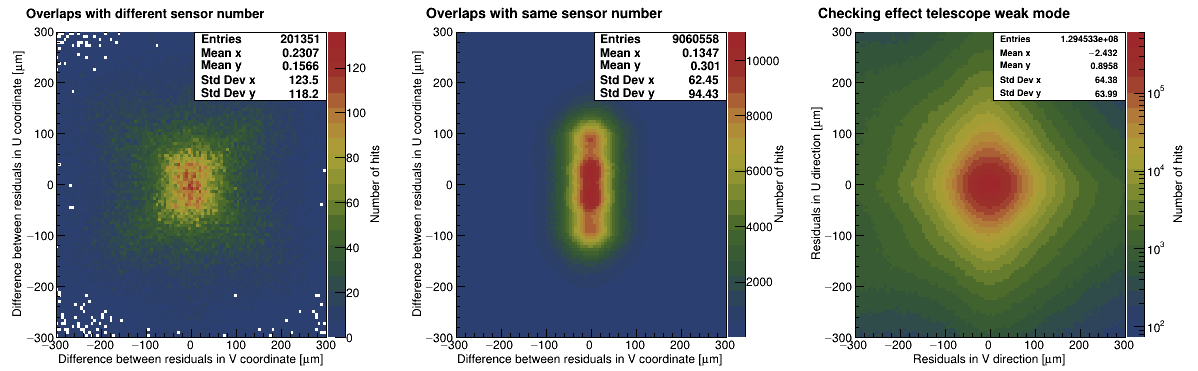}    &
 \includegraphics[width=35mm, trim={144mm, 0mm, 144mm, 8mm}, clip]{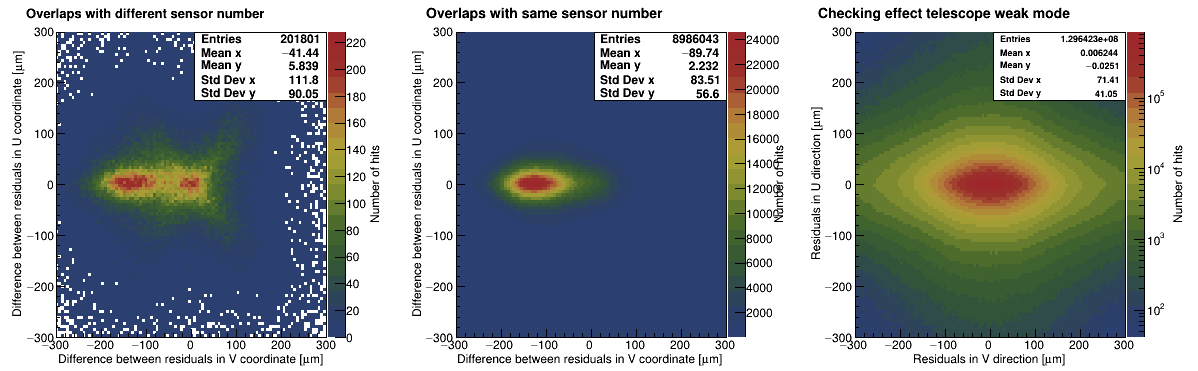} \tabularnewline \hline
\textbf{Curl} & \textbf{Z expansion} \tabularnewline
 \includegraphics[width=35mm, trim={144mm, 0mm, 144mm, 8mm}, clip]{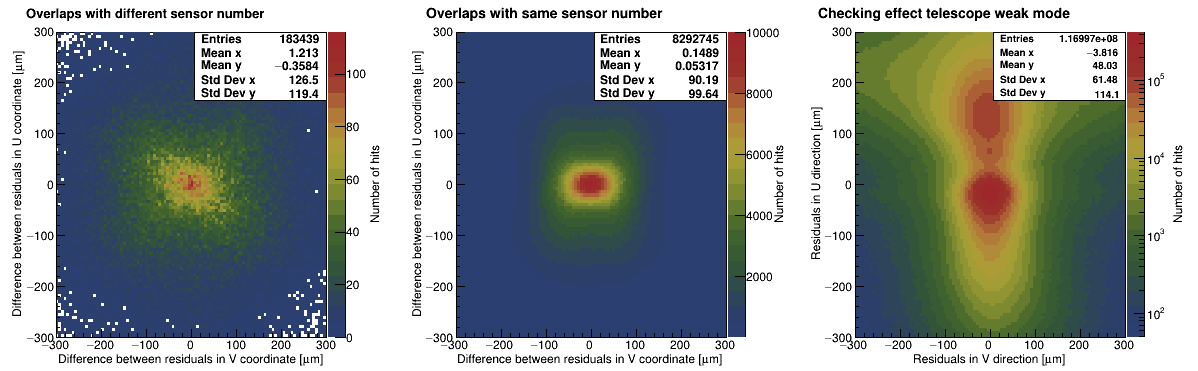}    &
 \includegraphics[width=35mm, trim={144mm, 0mm, 144mm, 8mm}, clip]{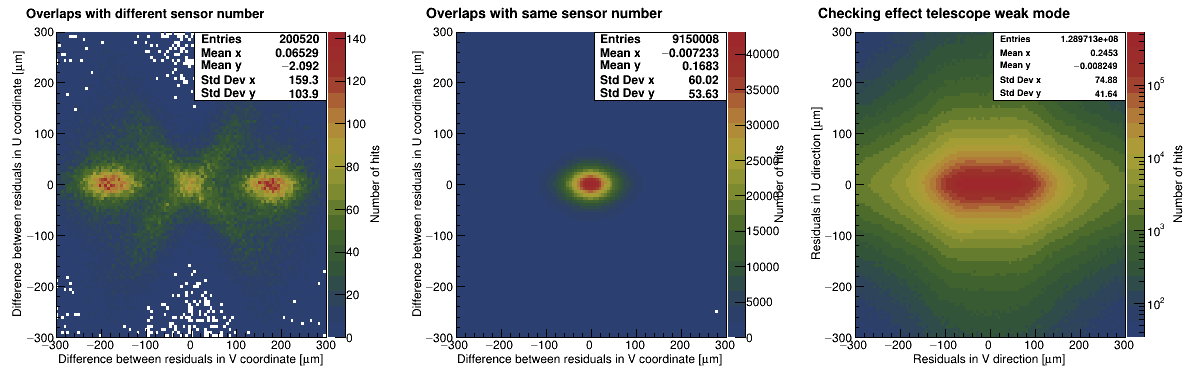} \tabularnewline
\end{tabular}
\caption{Comparison of Monte Carlo results for nominal geometry and $\chi^2$ invariant modes using overlapping hits with same position of sensors in ladders for cosmic dataset}
\label{fig:Cosmic}
\end{figure}

\begin{figure}[h]
\centering
\begin{tabular}{c|c} 
\textbf{Nominal} & \textbf{Clamshell} \tabularnewline
 \includegraphics[width=35mm, trim={144mm, 0mm, 144mm, 8mm}, clip]{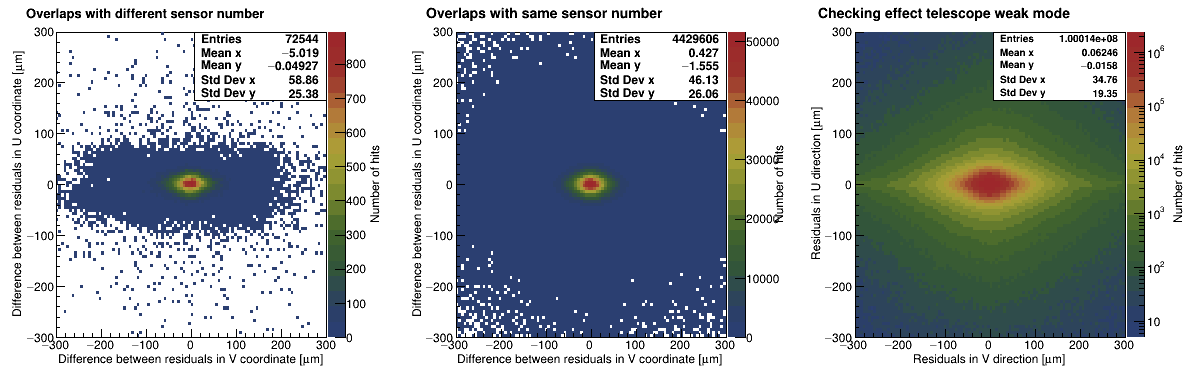}    &
 \includegraphics[width=35mm, trim={144mm, 0mm, 144mm, 8mm}, clip]{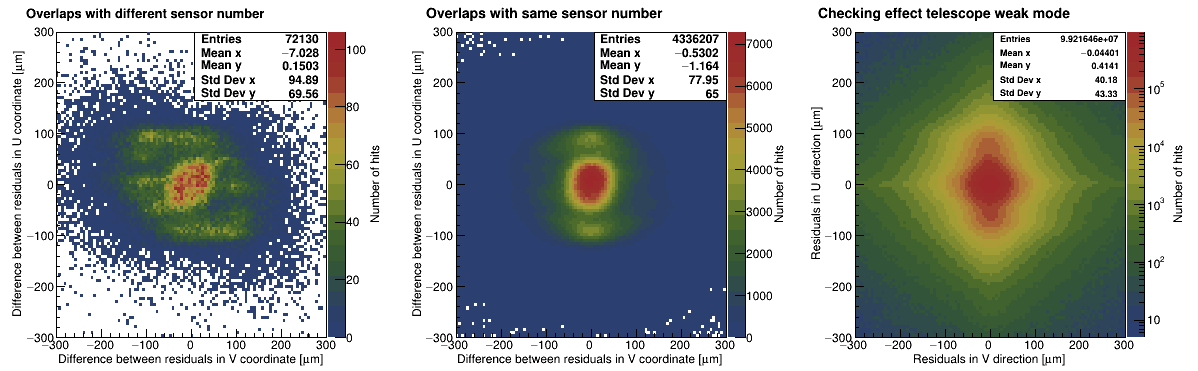} \tabularnewline \hline
\textbf{Radial expansion} & \textbf{Twist} \tabularnewline
 \includegraphics[width=35mm, trim={144mm, 0mm, 144mm, 8mm}, clip]{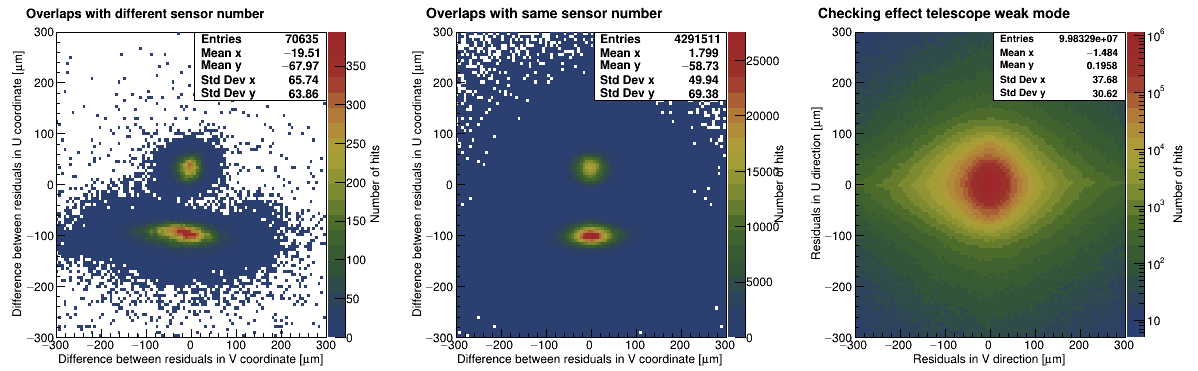}    &
 \includegraphics[width=35mm, trim={144mm, 0mm, 144mm, 8mm}, clip]{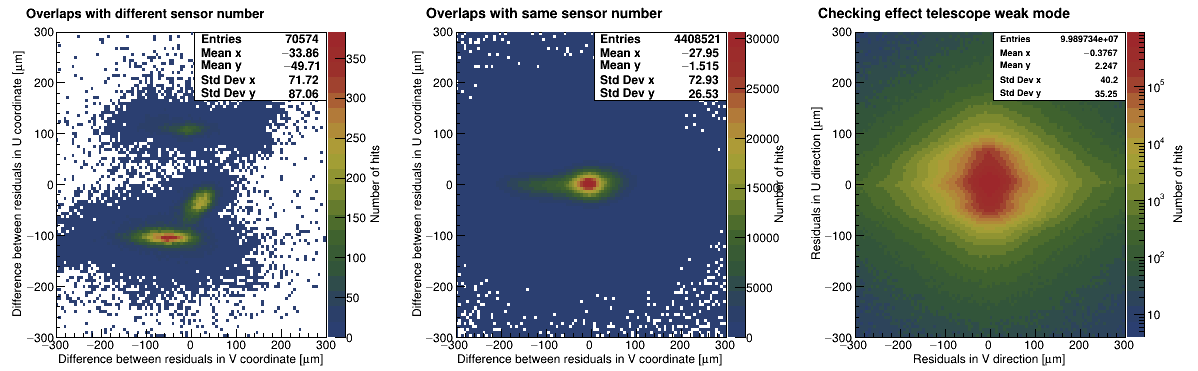} \tabularnewline \hline
\textbf{Elliptical expansion} & \textbf{Telescope} \tabularnewline
 \includegraphics[width=35mm, trim={144mm, 0mm, 144mm, 8mm}, clip]{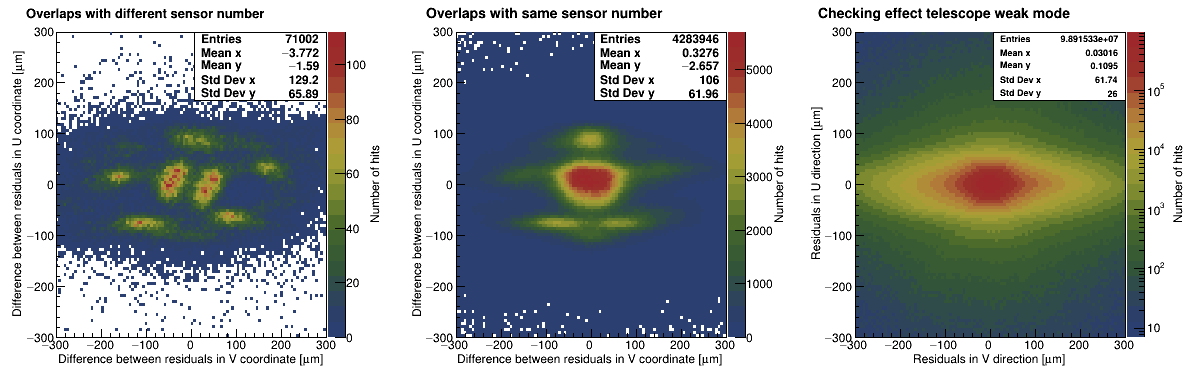}    &
 \includegraphics[width=35mm, trim={144mm, 0mm, 144mm, 8mm}, clip]{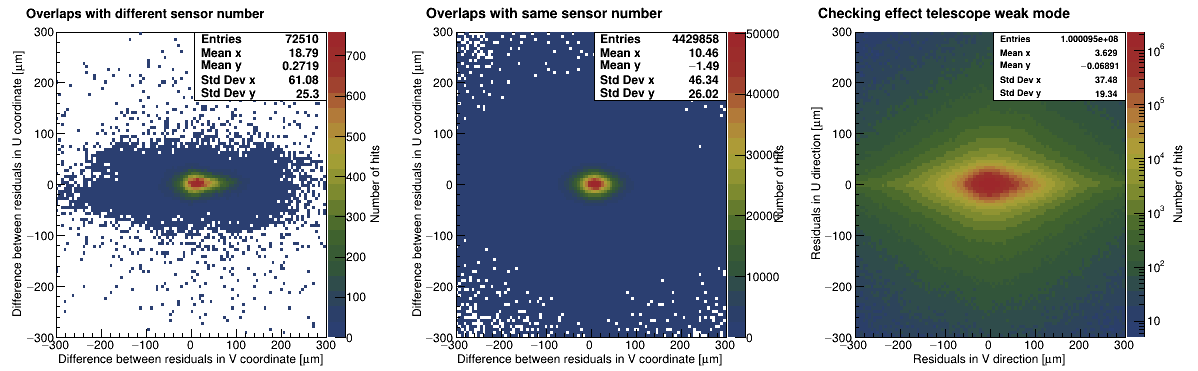} \tabularnewline \hline
\textbf{Bowing} & \textbf{Skew} \tabularnewline
 \includegraphics[width=35mm, trim={144mm, 0mm, 144mm, 8mm}, clip]{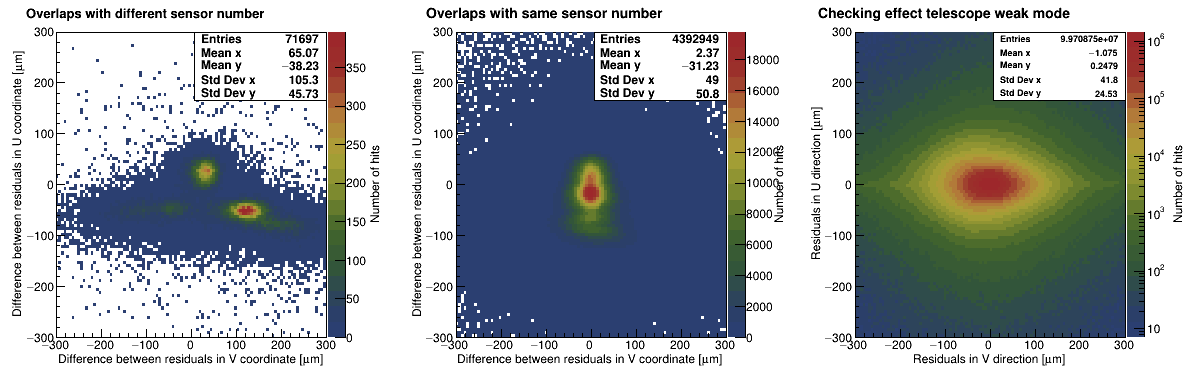}    &
 \includegraphics[width=35mm, trim={144mm, 0mm, 144mm, 8mm}, clip]{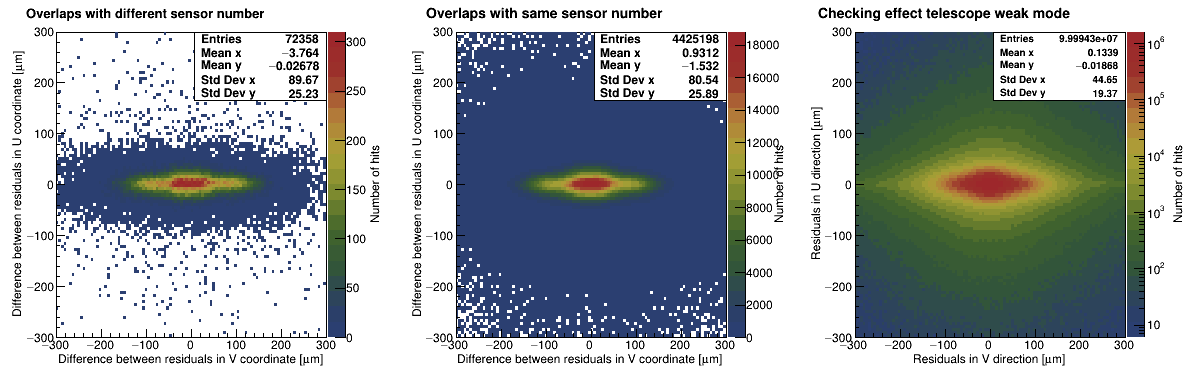} \tabularnewline \hline
\textbf{Curl} & \textbf{Z expansion} \tabularnewline
 \includegraphics[width=35mm, trim={144mm, 0mm, 144mm, 8mm}, clip]{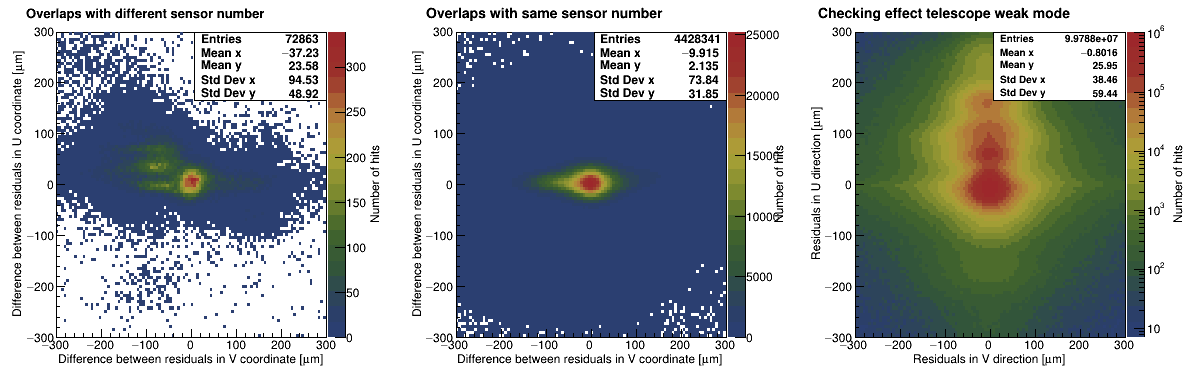}    &
 \includegraphics[width=35mm, trim={144mm, 0mm, 144mm, 8mm}, clip]{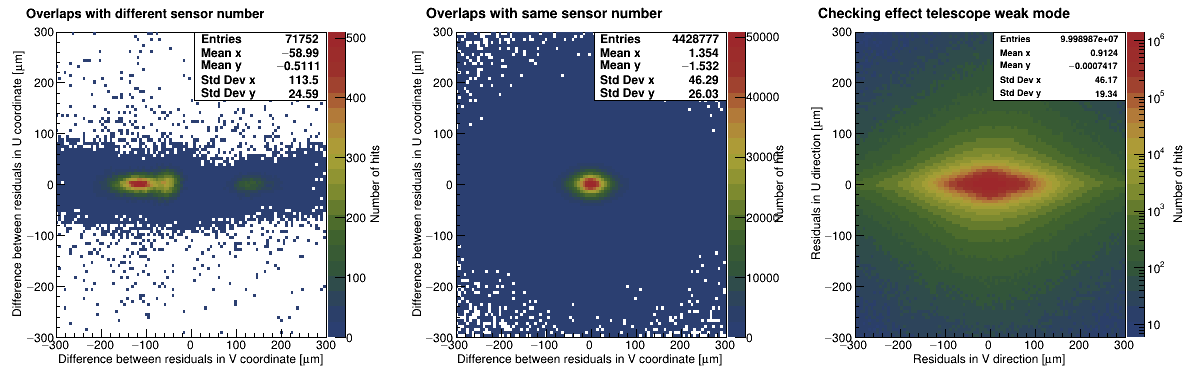} \tabularnewline
\end{tabular}
\caption{Comparison of Monte Carlo results for nominal geometry and weak modes using overlapping hits with same position of sensors in ladders for dimuon dataset}
\label{fig:Dimuons}
\end{figure}

\noindent
For clarifying properties and testing usefulness of monitoring tool is necessary provide Monte Carlo studies. The studies was divided to several directions: to universality assessment of monitoring tool and checking possibility identifying the most number of weak modes. The first property was analysed using different datasets: cosmic rays (muons) passing vertex detector without using magnetic field and pair of muons from $e^- + e^+ \to \mu^- + \mu^+$. Tracks from each of these datasets have different properties and effects to our studies are not negligible. Cosmic tracks are straight lines passing through vertex detector. An event is composed by one single track and his momentum range is from tens of MeV to TeV. Pairs of muons with specific momenta can be observed during electron-positron collisions. For both cases multiple scattering effect of tracks in overlapping area can be neglected because of high momentum of particles. \\

\noindent
With comparison collisions and cosmic datasets one was observed different distributions for same physical variables. Example of difference between datasets can be explained in figure~\ref{fig:Cosmic} and figure~\ref{fig:Dimuons} using overlapping hits with same position of sensors in ladders, there are comparing different geometry scenarios. Using more frequent type of overlapping hits many of $\chi^2$ invariant modes can be recognized from nominal geometry. However differences between nominal, telescope and $z$ expansion geometry are negligible and one can not recognized between them in both of datasets. \\

\begin{figure}[h]
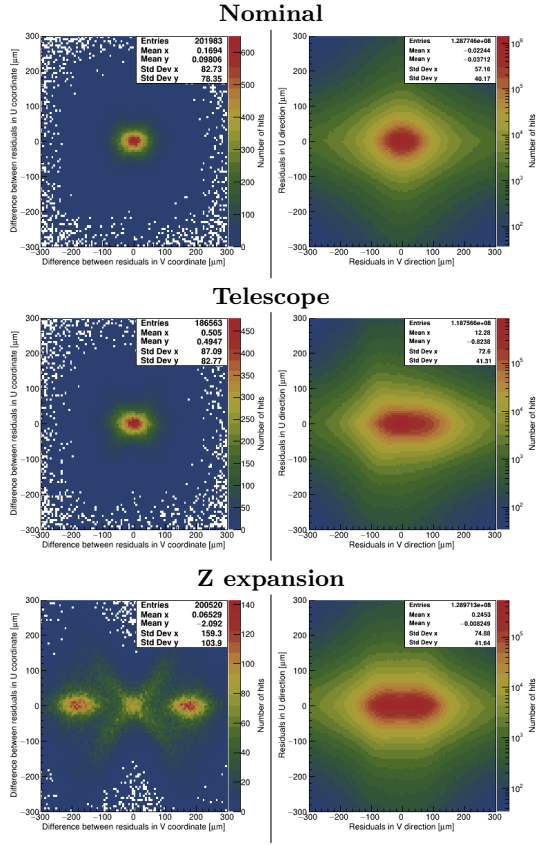

\centering
\begin{tabular}{c|c} 
\multicolumn{2}{c}{\textbf{Nominal}} \tabularnewline
 \includegraphics[width=35mm, trim={0mm, 0mm, 288mm, 8mm}, clip]{cosmic_Nominal.png} &
 \includegraphics[width=35mm, trim={285mm, 0mm, 3mm, 8mm}, clip]{cosmic_Nominal.png} \tabularnewline
\multicolumn{2}{c}{\textbf{Telescope}} \tabularnewline
 \includegraphics[width=35mm, trim={0mm, 0mm, 288mm, 8mm}, clip]{cosmic_Telescope.png} &
 \includegraphics[width=35mm, trim={285mm, 0mm, 3mm, 8mm}, clip]{cosmic_Telescope.png} \tabularnewline
\multicolumn{2}{c}{\textbf{Z expansion}} \tabularnewline
 \includegraphics[width=35mm, trim={0mm, 0mm, 288mm, 8mm}, clip]{cosmic_ZExpansion.png} &
 \includegraphics[width=35mm, trim={285mm, 0mm, 3mm, 8mm}, clip]{cosmic_ZExpansion.png} \tabularnewline
\end{tabular}
\caption{Comparison of Monte Carlo results for nominal telescope and $z$ expansion geometry using overlapping hits with different position of sensors in ladders for cosmic dataset}
\label{fig:CosmicZ}
\end{figure}

\begin{figure}[h]
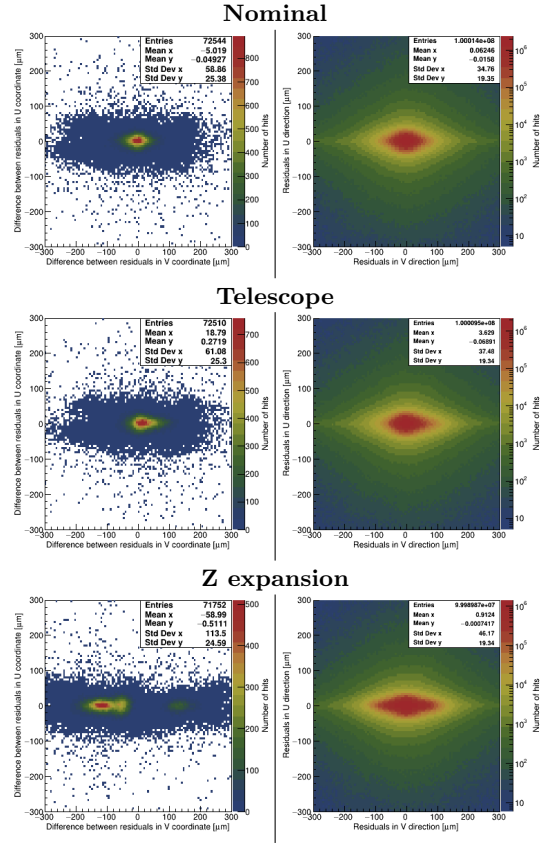

\centering
\begin{tabular}{c|c} 
\multicolumn{2}{c}{\textbf{Nominal}} \tabularnewline
 \includegraphics[width=35mm, trim={0mm, 0mm, 288mm, 8mm}, clip]{dimuons_Nominal.png} &
 \includegraphics[width=35mm, trim={285mm, 0mm, 3mm, 8mm}, clip]{dimuons_Nominal.png} \tabularnewline
\multicolumn{2}{c}{\textbf{Telescope}} \tabularnewline
 \includegraphics[width=35mm, trim={0mm, 0mm, 288mm, 8mm}, clip]{dimuons_Telescope.png} &
 \includegraphics[width=35mm, trim={285mm, 0mm, 3mm, 8mm}, clip]{dimuons_Telescope.png} \tabularnewline
\multicolumn{2}{c}{\textbf{Z expansion}} \tabularnewline
 \includegraphics[width=35mm, trim={0mm, 0mm, 288mm, 8mm}, clip]{dimuons_ZExpansion.png} &
 \includegraphics[width=35mm, trim={285mm, 0mm, 3mm, 8mm}, clip]{dimuons_ZExpansion.png} \tabularnewline
\end{tabular}
\caption{Comparison of Monte Carlo results for nominal telescope and $z$ expansion geometry using overlapping hits with different position of sensors in ladders for dimuon dataset}
\label{fig:DimuonsZ}
\end{figure}

\noindent
To recognized changes in $z$ it is useful to study overlapping hits with different position of sensors in ladders or residual distribution for non-overlapping hits. According figure~\ref{fig:CosmicZ} and figure~\ref{fig:DimuonsZ} we can recognized $z$ expansion from nominal geometry in both datasets. Recognition telescope scenario is not very clear, but possible. \\

\noindent 
Clarifying correct geometry of the Belle II vertex detector can be done in two steps: quick check using cosmic data and precise check using $e^+ + e^- \to \mu^+ + \mu^-$ dataset. According Monte Carlo studies high statistics of reconstructed tracks in needed, at least one million of cosmic muons passing vertex detector for quick check. After collecting enough collision data with stable alignment conditions we are able validate vertex detector geometry more precisely. \\

\begin{figure}[h]
\centering
\begin{tabular}{c|c}
\rule{0pt}{2ex}
\textbf{February 2019} & \textbf{April 2019} \tabularnewline
 \includegraphics[width=35mm, trim={144mm, 0mm, 144mm, 8mm}, clip]{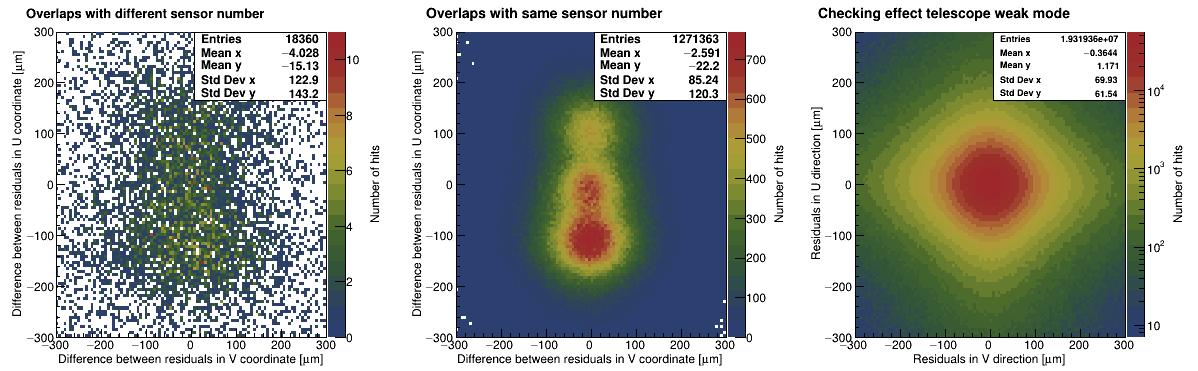} &
 \includegraphics[width=35mm, trim={144mm, 0mm, 144mm, 8mm}, clip]{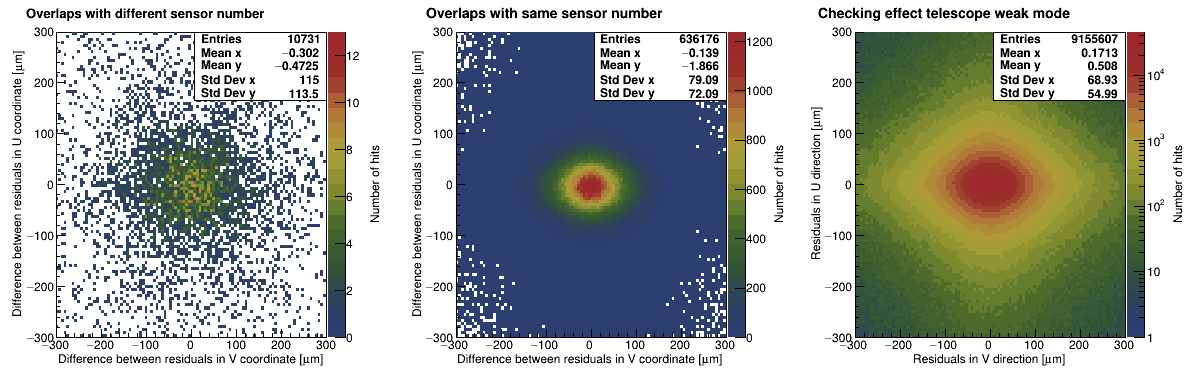} \tabularnewline
 \includegraphics[width=35mm, trim={0mm, 0mm, 288mm, 8mm}, clip]{OverlapStudyExp5.png} &
 \includegraphics[width=35mm, trim={0mm, 0mm, 288mm, 8mm}, clip]{OverlapStudyExp7.png} \tabularnewline 
 \includegraphics[width=35mm, trim={285mm, 0mm, 3mm, 8mm}, clip]{OverlapStudyExp5.png} &
 \includegraphics[width=35mm, trim={285mm, 0mm, 3mm, 8mm}, clip]{OverlapStudyExp7.png} \tabularnewline
\end{tabular}
\caption{Identifying $\chi^2$ invariant modes of Belle II vertex detector using cosmic data taken in February (left) and April (right) 2019: The rows present residual differences for overlaps with same (top) or different (center) positions of sensors in ladders and residual distributions for hits from non-overlapping area (bottom).}
\label{fig:Data}
\end{figure}

\section{Cosmic data and $\chi^2$ invariant modes}
\noindent 
Since installation of the Belle II vertex detector the detector was started collecting cosmic tracks. In first period (in February 2019) we collected about 3.4 millions of cosmic muons passing vertex detector. Analysing this data we observed result in figure~\ref{fig:Data}. Validation plot for overlapping hits with same sensor positions in layers present strange result. The pattern, is looked like ''snowman'', presents unexpected residual difference distribution. One of advantage developed tool is possibility to look at residual difference distributions as function of layers, ladders or sensors. At detailed look we found different contributions from parts of vertex detector to final result: pixel detector part contribute to central ball of ''snowman'', slanted sensors of strip detector contribute to central ball too and barrel part of strip detector contribute to top and bottom ball of snowman. The contribution of barrel part has similar behaviour as radial expansion of vertex detector in Monte Carlo studies. \\

\noindent
After investigation in reconstruction software and realigning vertex detector the study using overlapping hits was providing again with different collection of data (April 2019). The result, there was used 4.8 million cosmic tracks, is shown in figure~\ref{fig:Data}. This data presents correct geometry of the Belle II vertex detector and reconstruction software. According the latest result we can prove, the data is not affected any systematic misalignment because of minimization algorithm used for alignment of vertex detector. 

\section{B lifetime and $B^0-\bar B^0$ mixing measurements}
\noindent
Monte Carlo studies present several ways to measure B lifetime and $B^0-\bar B^0$ mixing using the Belle II detector. Analyses are based on hadronic and semileptonic B decays. Technique using hadronic B decays is standard method for B factories. The technique will use Full Event Interpretation~\cite{FEI}, vertex reconstruction and flavor tagging software of the Belle II detector~\cite{FlavorTagger}. Otherwise semileptonic technique is not standard, but simpler than hadronic technique. The technique is based on selection high momentum leptons, which are pure with high branching fraction of $B \to X l \nu$ for $e, \mu$. The B vertices can be found as intersection of lepton tracks with beamspot. \\

\noindent
Semileptonic B decays technique can use several approaches for measurement B lifetime and $B^0-\bar B^0$ mixing: fully inclusive, full reconstruction of $B \to D^{(*)} l \nu$ and partial reconstructed $B \to D^{*-} (\to \bar{D}^0 \pi^-_{soft}) l^+ \nu$. Full inclusive approach is based on two lepton event selection, what is very probable in the Belle II detector. However, identifying between charged and neutral B mesons is impossible in this case and modelling of background should not be trivial. Using full reconstruction of $B \to D^{(*)} l \nu$ approach we are able to recognize between charged and neutral B mesons, background of this process is more known and it can be controlled. However it has lower efficiencies than other approaches. The last approach, partial reconstruction of $B \to D^{*-} (\to \bar{D}^0 \pi^-_{soft}) l^+ \nu$, can be used too. The $D^*$ momentum can be interred from $\pi_{soft}$ momentum, but the background is high for this approach. 

\begin{figure}[h]
\centering
\includegraphics[width=80mm, trim={0mm, 0mm, 0mm, 0mm}, clip]{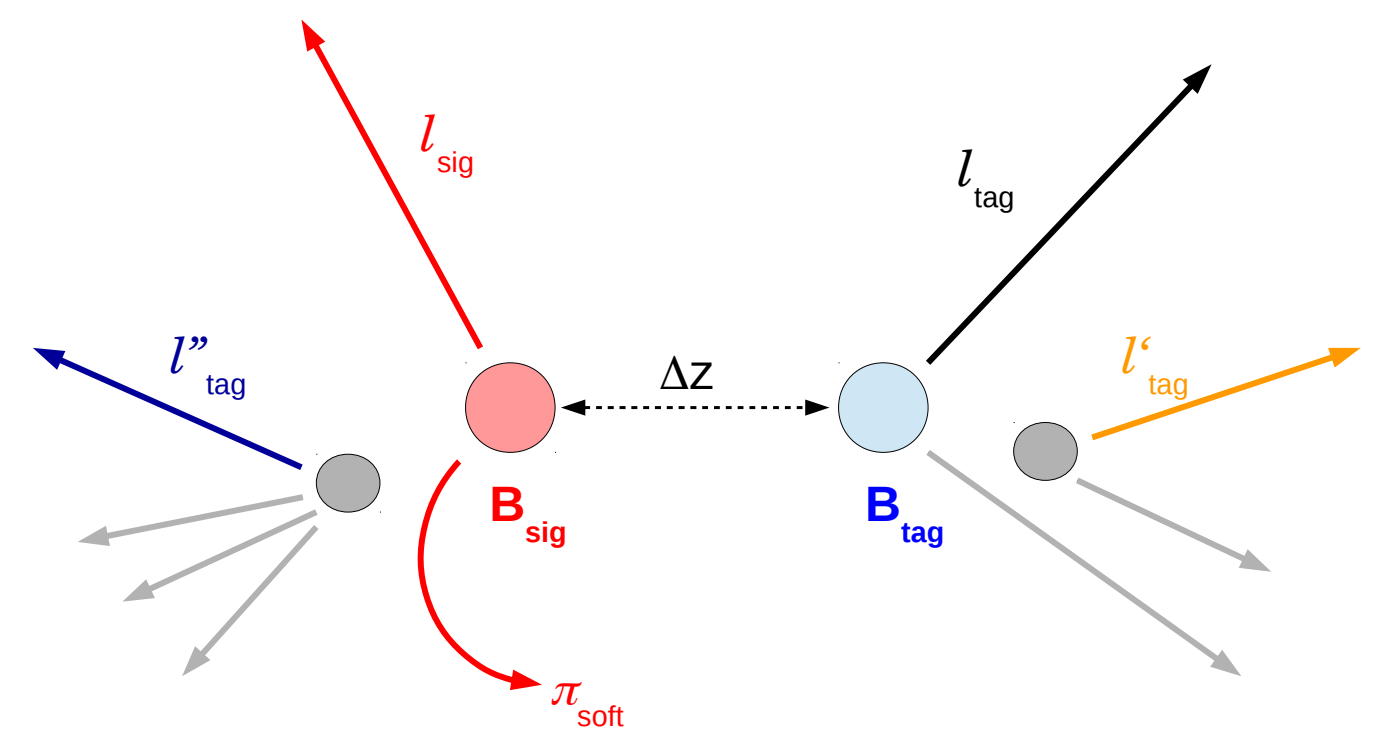}
\caption{Measurement of distance between the signal side B candidate (red circle) and the tag side B (blue circle) using two lepton approach. Open charmed particles (grey circles) are displaced from mother B mesons.} 
\label{fig:SchematicPhysics}
\end{figure}

\noindent
The approach using two leptons can be describe using figure~\ref{fig:SchematicPhysics}. The signal side B candidate $\mathrm{B_{sig}}$ (red light circle) is founded using ($\pi_{soft}, l$) pair (red arrows). $\bar{D}^0$ meson (light gray circle) is not reconstructed. If $\mathrm{B_{sig}}$ is  correctly reconstructed, $l_{tag}$ can come from:
\begin{enumerate}
\item[I)]{$l_{tag}$ (black arrow) is identified correctly from tag side B candidate $B_{tag}$. This selected event is signal event and spacial separation between two B decay vertices $\Delta z$ is unbiased.}
\item[II)]{$l'_{tag}$ (orange arrow) is identified correctly from tag side B candidate $B_{tag}$, but an origin of track is open charm meson with displaced vertex. $\Delta z$ is biased by non-negligible life time of open charm mesons.}
\item[III)]{$l''_{tag}$ (blue arrow) is identified from signal side B candidate $B_{sig}$. Measurement $\Delta z$ is not difference between two B mesons, but measurement of convolution $D^0$ lifetime convoluted with resolution function.}
\end{enumerate}

\begin{figure*}[t]
\centering
\includegraphics[width=135mm]{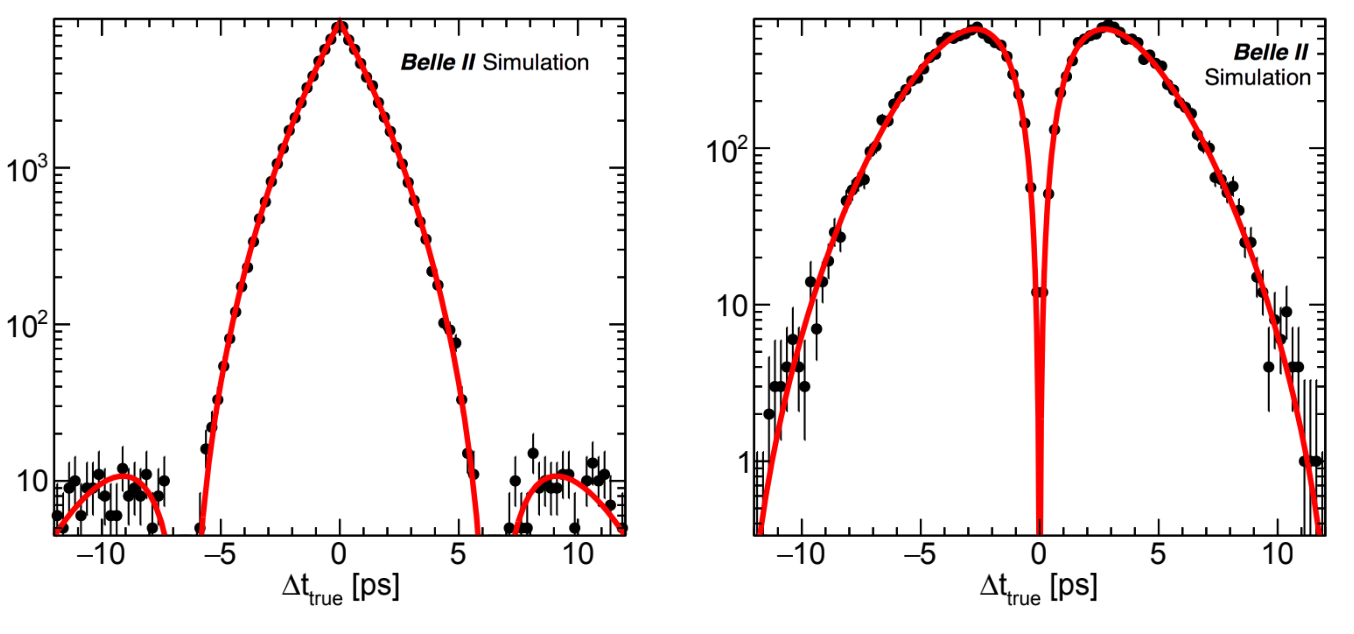}
\caption{A fit to the $\Delta t_{true}$ distribution for unmixed (left) and unmixed (right) simulated samples} \label{fig:true}
\end{figure*}

\begin{figure*}[t]
\centering
\includegraphics[width=135mm]{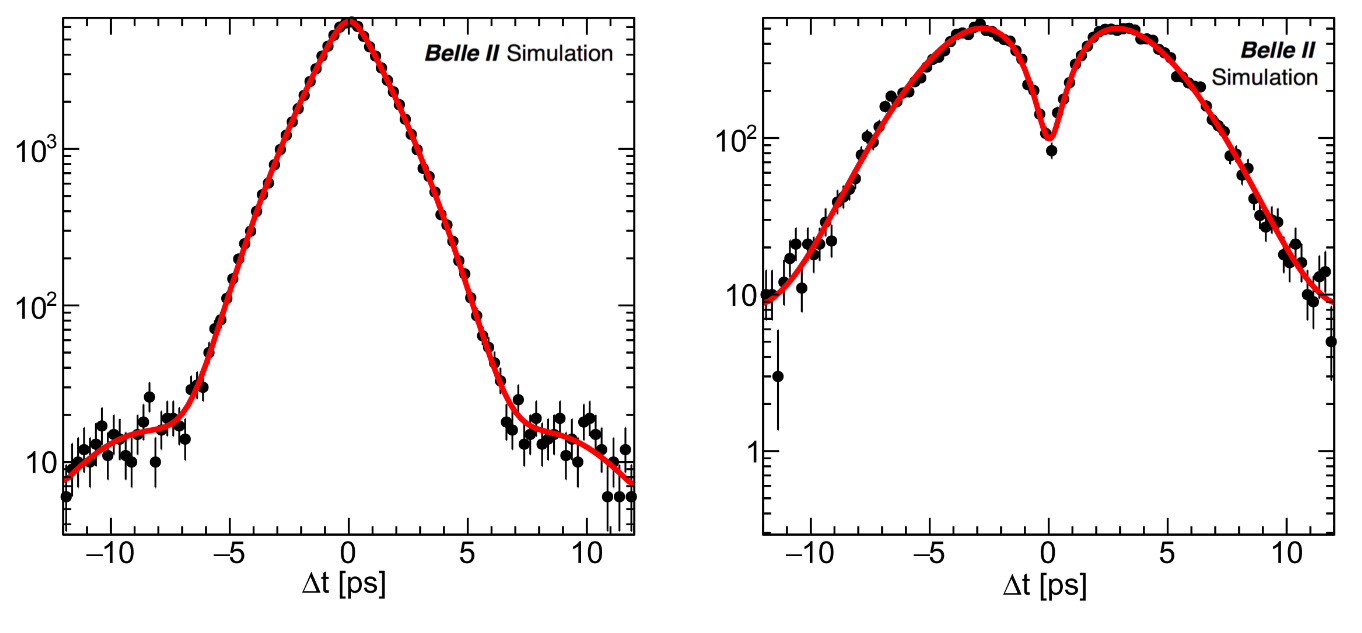}
\caption{A fit to the $\Delta t$ distribution for unmixed (left) and unmixed (right) simulated samples} \label{fig:reco}
\end{figure*}

\noindent
All possible cases have negligible affect to final $\Delta z$ distributions. To determine $\Delta t$ resolution is used transformation $\Delta t = \Delta z / (\theta_{boost} \beta \gamma c)$, where $\beta \gamma$ is Lorentz boost of the Center of Mass System (CMS), $c$ is speed of light and $\theta_{boost} \ \sim \ \mathrm{0.15 \ rad}$ is angle between boost axis and z axis of global reference. \\

\noindent
The Monte Carlo studies focused on $\Delta t$ resolution were provided. The standard probability density functions for B lifetime and $B^0-\bar B^0$ mixing measurement is used:
\begin{equation}
P^{\pm}(\Delta t) = \frac{e^{-|\Delta t|/\tau_{B^0}}}{\tau_{B^0}}(1 \pm \cos{\Delta m \Delta t}), \label{eq:physics}
\end{equation}
where function $P^+(\Delta t)$ corresponds with unmixed case $B^0\bar{B}^0 \to B^0\bar{B}^0$ and function $P^-(\Delta t)$ with mixed case $B^0\bar{B}^0 \to B^0B^0/\bar{B}^0\bar{B}^0$. In figure~\ref{fig:true} the functions are fitted using generated true distributions for unmixed and mixed cases. To take account measurement using real detector the resolution functions $\mathcal{R}^{\pm}$ should be defined as sum three Gaussians $\mathcal{G}$:
\begin{equation}
\mathcal{R}^{\pm} = f_1^{\pm}\mathcal{G}(\mu_1^{\pm}, \sigma_1^{\pm}) + f_2^{\pm}\mathcal{G}(\mu_2^{\pm}, \sigma_2^{\pm}) + f_3^{\pm}\mathcal{G}(\mu_3^{\pm}, \sigma_3^{\pm}). \label{eq:resolution}
\end{equation}

\noindent
A convolutions between probability density functions (equation \ref{eq:physics}) and resolution functions (equation \ref{eq:resolution}) fitted on simulated data for unmixed and mixed samples can be founded in figure~\ref{fig:reco}. 

\section{Conclusion}
\noindent
The Belle II vertex detector is installed and it is providing precise measurement. Applied alignment procedure it obtains alignment corrections. New tool for monitoring misalignment systematics was developed. His universality and usefulness was tested in Monte Carlo studies. The both datasets present sensitivity to validate correct geometry of vertex detector. According cosmic studies vertex detector is not affected by any $\chi^2$ invariant mode or combination of them. After collecting enough statistics the Belle II detector is able to provides B lifetime and $B^0-\bar B^0$ mixing measurements. 

\bigskip 

\begin{thebibliography}{9}   
\bibitem{DesignReport} {Z. Dolezal, S. Uno et al.} Technical Design Report, arxiv: 1011.0352v1

\bibitem{KandraThesis} {J. Kandra} Diploma thesis, Charles University (2016) 

\bibitem{FPCP2017} {T. Bilka, J. Kandra} Alignment and physics performance of the Belle II vertex detector, PoS FPCP2017 (2017) 053

\bibitem{LHCAlignment} {S. Blusk et al.} First LHC Detector Alignment Workshop, CERN-2007-004 (2007)

\bibitem{HeinemannThesis} {F. Heinemann} DPhil. Thesis, Oxford University (2007)

\bibitem{FEI} {Keck, T. et al.} Comput Softw Big Sci (2019) 3: 6. https://doi.org/10.1007/s41781-019-0021-8

\bibitem{FlavorTagger} {Moritz J. Gelb} Diploma thesis, Karlsruhe Institute of Technology (2015)

\end{thebibliography}

\end{document}